\documentclass[manuscript, screen]{acmart}
\AtBeginDocument{%
  }

\setcopyright{acmlicensed}
\copyrightyear{2025}
\acmYear{2025}
\acmDOI{XXXXXXX.XXXXXXX}

\usepackage{mdframed}
\usepackage{tcolorbox}
\usepackage{blindtext}
\usepackage[linesnumbered,boxed]{algorithm2e}
\usepackage{comment}
\usepackage{caption}
\usepackage{tabularx}
\usepackage{booktabs}
\usepackage{tikz}
\usepackage{graphicx}
\usepackage{subcaption}
\usepackage{array}
\usepackage{needspace}
\usepackage{tabularx}
\usepackage[utf8]{inputenc}
\usepackage{longtable}
\usepackage{caption}
\usepackage{listings}
\usepackage{float}
\usepackage{xcolor}

\definecolor{keycolor}{rgb}{0.0, 0.0, 0.6}
\definecolor{stringcolor}{rgb}{0.6, 0.2, 0.2}
\definecolor{numbercolor}{rgb}{0.0, 0.5, 0.0}
\definecolor{bracecolor}{rgb}{0.3, 0.3, 0.3}
\definecolor{backgroundcolor}{rgb}{0.97, 0.97, 0.97}
\definecolor{darkgreen}{rgb}{0.0, 0.5, 0.0}

\lstdefinelanguage{json}{
  basicstyle=\ttfamily\small,
  showstringspaces=false,
  breaklines=true,
  literate=
   *{:}{{{\color{bracecolor}:}}}{1}
    {,}{{{\color{bracecolor},}}}{1}
    {\{}{{{\color{bracecolor}\{}}}{1}
    {\}}{{{\color{bracecolor}\}}}}{1}
    {[}{{{\color{bracecolor}[}}}{1}
    {]}{{{\color{bracecolor}]}}}{1},
  morestring=[b]",
  stringstyle=\color{stringcolor},
  commentstyle=\color{gray},
  morecomment=[l]{//},
  keywordstyle=\color{keycolor},
  alsoletter={0123456789},
}

\lstdefinestyle{sqlstyle}{
    language=SQL,
    basicstyle=\ttfamily\small,
    keywordstyle=\color{darkgreen}\bfseries,
    commentstyle=\color{gray}\itshape,
    stringstyle=\color{orange},
    frame=single,
    breaklines=true,
    showstringspaces=false,
    tabsize=2,
    morekeywords={SERIAL, PRIMARY, KEY, REFERENCES, NOT, NULL, CONSTRAINT, UNIQUE, DEFAULT, TIMESTAMP, CHECK, IN, ON, DELETE, CASCADE}
}

\SetAlCapNameFnt{\normalfont\small}
\SetAlCapFnt{\normalfont\small}
\SetAlCapNameSty{}
\SetAlCapSty{}

\renewcommand{\arraystretch}{1.3}

\begin{document}

\definecolor{minipagecolor}{rgb}{0.9,0.9,1} 
\definecolor{codepurple}{rgb}{0.58,0,0.82}
\definecolor{codegreen}{rgb}{0,0.6,0}
\definecolor{lightgray}{gray}{0.9}
\definecolor{limegreen}{rgb}{.725, 1, .40}
\definecolor{lightred}{rgb}{.973, .514, .475}

\newcommand{\Var}{\textup}
\newcommand{\algorithmicbreak}{\textbf{break}}
\newcommand{\BREAK}{\STATE \algorithmicbreak}
\SetKwInput{KwRequire}{Require}
\SetNlSty{}{}{:}
\setlength{\textfloatsep}{0pt}

\newcolumntype{L}{>{\raggedright\arraybackslash}X}

\title{VerilogDB: The Largest, Highest-Quality Dataset with a Preprocessing Framework for LLM-based RTL Generation}



\author{Paul E. Calzada}
\email{paul.calzada@ufl.edu}
\orcid{0000-0002-5001-6321}
\author{Zahin Ibnat}
\email{ibnatz16@ufl.edu}
\orcid{0000-0001-5664-4428}
\author{Tanvir Rahman}
\email{tanvir.rahman@ufl.edu}
\author{Kamal Kandula}
\email{kamalkandula@ufl.edu}
\author{Danyu Lu}
\email{danyulu@ufl.edu}
\author{Sujan Kumar Saha}
\orcid{0009-0009-5311-9367}
\email{sujansaha@ufl.edu}
\author{Farimah Farahmandi}
\orcid{0000-0003-1535-0938}
\email{farimah@ece.ufl.edu}
\author{Mark Tehranipoor}
\orcid{0000-0003-4699-3231}
\email{tehranipoor@ece.ufl.edu}
\affiliation{%
  \institution{University of Florida}
  \city{Gainesville}
  \state{Florida}
  \country{USA}
}


\renewcommand{\shortauthors}{Calzada et al.}

\begin{abstract}
Large Language Models (LLMs) are gaining popularity for hardware design automation, particularly through Register Transfer Level (RTL) code generation. In this work, we examine the current literature on RTL generation using LLMs and identify key requirements for training and fine-tuning datasets. We construct a robust Verilog dataset through an automated three-pronged process involving database (DB) creation and management with PostgreSQL \cite{postgresql}, data collection from code hosting sites like OpenCores and GitHub, and data preprocessing to verify the codes' syntax, run logic synthesis, and extract relevant module metadata. We implement a scalable and efficient DB infrastructure to support analysis and detail our preprocessing pipeline to enforce high-quality data before DB insertion. The resulting dataset comprises 20,392 Verilog samples, 751 MB of Verilog code data, which is the largest high-quality Verilog dataset for LLM fine-tuning to our knowledge. We further evaluate the dataset, address associated challenges, and explore potential applications for future research and development in LLM-based hardware generation.

\end{abstract}




\maketitle

\section{Introduction}


The push for quicker time-to-market (TTM) in the semiconductor industry has heightened the necessity for automation in hardware design, aiming to enhance power, performance, and area (PPA) metrics. This automation of hardware design, particularly for System-on-Chip (SoC) development, is rapidly evolving with the emergence of LLMs. The generative artificial intelligence (GenAI) models, initially developed for natural language and high-level programming tasks, are increasingly being adapted to support code generation in hardware description languages (HDLs) such as Verilog. However, for LLMs to be effective in this domain, they require clean, diverse, and well-structured training datasets that reflect real-world RTL design patterns and constraints.

Recent efforts such as \textit{VeriGen}~\cite{thakur2024verigen} and \textit{VerilogEval}~\cite{liu2023verilogeval} have highlighted both the promise and limitations of applying LLMs to Verilog generation. These studies demonstrate that while LLMs can produce syntactically valid Verilog, their ability to generalize and generate semantically meaningful logic is heavily dependent on the quality and diversity of the training data. For instance, \textit{VeriGen} fine-tuned a code generation model using GitHub-sourced Verilog code and textbook examples, but the dataset required extensive manual filtering to remove simulation-specific or non-synthesizable files. Similarly, \textit{VerilogEval} established a standardized evaluation suite for Verilog LLMs but noted performance degradation in modules that deviated from the model’s perceived patterns.

The absence of high-quality curated Verilog datasets remains a significant area for improvement. Verilog code available in public repositories is not particularly reliable when used for data-driven research~\cite{liu2023verilogeval}. While Verilog is a standardized HDL, the code found in open-source platforms such as GitHub or OpenCores varies significantly in quality, style, and completeness. Projects often differ in indentation, naming conventions, commenting practices, and overall formatting. Some modules are clearly defined and self-contained, while others are deeply nested within larger designs or split across multiple files with ambiguous dependencies. These inconsistencies cause problems both in manual inspection and in automated parsing~\cite{liu2023verilogeval}. Moreover, Verilog frequently includes preprocessing directives, macros, and vendor-specific syntax, which cannot be used for standardization.

For LLMs to effectively generate or understand Verilog code, high-quality and consistently structured training data is needed. Unlike high-level programming languages, where syntactic errors may be caught and corrected during runtime or compilation, errors in HDL can propagate through synthesis and further to the bitstream that will ultimately cause major problems in hardware. These errors can exist in major sources of data and as a result, LLMs trained on poorly curated HDL corpora are prone to generating incorrect or logically incoherent outputs, which is not only unhelpful but potentially dangerous in security-critical systems~\cite{wang2025largelanguagemodelverilog}. This proves the need to increase dataset quality in LLM-driven hardware design generation. 


To address the aforementioned issues, we develop \textit{VerilogDB}: a scalable and automated pipeline to collect, sanitize, and structure Verilog RTL data from diverse open-source locations. Such a DB allows for the storage and indexing of HDL samples in a way that supports scalability, filtering, and metadata tagging~\cite{allam2024rtlrepobenchmarkevaluatingllms}. Our primary objective is to produce a high-quality dataset suitable for LLM fine-tuning in the context of IP and SoC design generation. The contributions of this work are as follows:

\begin{itemize}
    \item We curate a large corpus of Verilog RTL files by scraping GitHub, OpenCores, and university course materials, while applying structured keyword filters to exclude simulation-only or library-based repositories. We used a robust filtering approach to exclude codes with non-synthesizable constructs, testbenches, and system-verilog codes. We are the first, to our knowledge, to include hierarchical designs in the dataset.
    \item We implement an extensive preprocessing pipeline that includes duplicate removal (using content-based hashing), syntax validation (using \texttt{iverilog \cite{iverilog}}), and logic synthesis checking (using \texttt{Yosys \cite{yosys}}). This flow is automated compared to other works that rely on manual sample preparation and validation. We are the first, to our knowledge, to include explicit logic synthesis checks in Verilog data cleaning.
    \item From the cleaned modules, we extract detailed metadata including module names, I/O definitions, inline comments, token counts, and short natural language descriptions into a JSON format to support instruction-based LLM training.
    \item We demonstrate the scalability of our pipeline, which processes over 30 GB of raw data and produces a clean, structured dataset consisting of 20,392 synthesizable Verilog modules, at a 751 MB data size, ready for downstream learning tasks. This dataset is the largest of its kind to our knowledge. We also analyze the dataset's module complexity and diversity and provide further insights of our work to LLM training for RTL generation.
\end{itemize}

Section \ref{section:background} discusses RTL-generation through LLMs and considerations for generated Verilog code, and explains the
current state-of-the-art in LLM-based RTL generators. Section \ref{section:verilogdb} presents each component of our methodology including data collection, preprocessing, and insertion into the DB. The next section provides an analysis of the resulting dataset while outlining its application to support LLM-based RTL generation and hardware design automation. The benefits and challenges of \textit{VerilogDB} are also discussed. Section \ref{section:conclusion} concludes the work.

\section{Background}
\label{section:background}




\subsection{LLM Fine-Tuning for RTL Generation}


\begin{figure}[htbp]
\centering
\includegraphics[width=\linewidth, trim=0 0 0 0, clip]{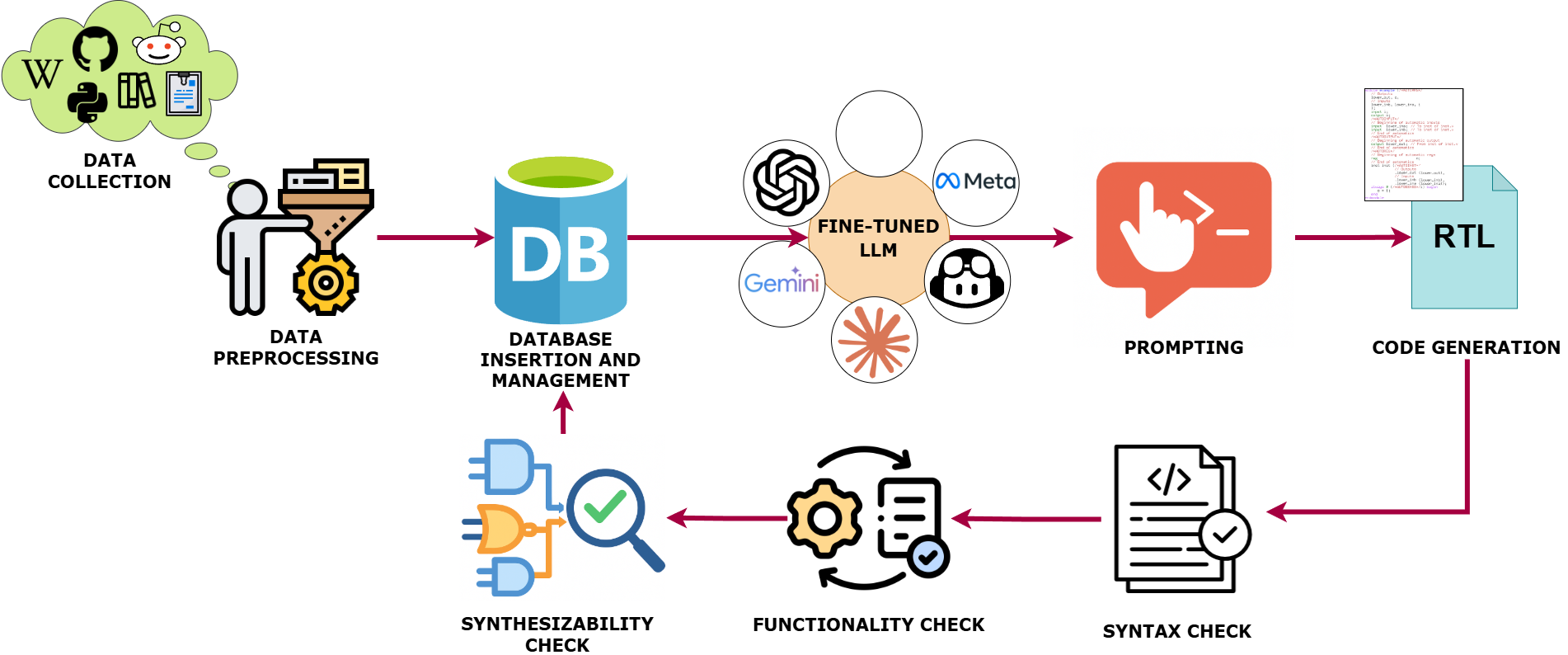}
\vspace{-8pt}
\caption{High-level flow for LLM fine-tuning process including iterative feedback for closed-loop fine-tuning.}
\vspace{6pt}
\label{fig:framework}
\end{figure}

Curated domain-specific datasets have served as the backbone for many innovative ML tasks from fine-tuning to retrieval augmented generation (RAG) pipelines. As shown in Figure~\ref{fig:framework}, developing a fine-tuned LLM for RTL generation must follow a meticulous workflow to optimize performance. The first step is the generation of a catered dataset of RTL sourced from diverse codebases which then undergoes rigorous preprocessing and is organized into a scalable database architecture. The sources and quality of data dictate the efficacy of LLM RTL generation, which is explored in this section with discussions on the quality of code in Section~\ref{subsec:background_code} and the subsequent results of leading LLM RTL code generation solutions in Section~\ref{subsec:background_prior}. Next, the stored data is used to fine-tune an open-source LLM, especially those already trained on a large corpus of code. Then, a procedural prompting approach can be deployed to generate an RTL code output set. These outputs are then evaluated by syntax, functionality via testbenches, and logic synthesis to gauge the efficacy of the model. The cleaned data is fed back into the database to further enhance the DB and overall model performance \cite{wu2025itertliterativeframeworkfinetuning}. The feedback data must also be preprocessed similarly to the initial data to enforce consistency. This meticulous process is not always followed through for a large corpus of data, and as a result, can cause the fine-tuned models or agents to generate inconsistent results.

\subsection{Considerations for LLM-Generated Verilog Code} \label{subsec:background_code}

The application of LLMs to hardware design tasks is a growing field, driven by the success of software code generation. While general-purpose code generation models like OpenAI's Codex~\cite{openai2025codex} and Code LLaMA~\cite{roziere2024code} show strong performances on objected-oriented programming languages (e.g., Python, C++), their capabilities in the hardware domain, particularly for Verilog RTL generation, remain underexplored and are likely limited because of the scarcity of hardware-specific training data. Moreover, hardware design has strict correctness constraints. Additionally, Verilog is a concurrent, timing-sensitive language where errors may not present themselves just in syntax and simulation. Verilog also includes both behavioral and structural modeling styles, vendor-specific extensions, and preprocessor macros. As a result, domain-specific fine-tuning is crucial for LLMs to understand RTL constructs, timing, and logic synthesis requirements. The dataset must consist of:

\begin{itemize}

    \item \textbf{Syntactically valid and synthesizable Verilog modules.} 
    \vspace{5pt}
    \begin{mdframed}[backgroundcolor=limegreen] 
    \begin{minipage}{\columnwidth}
    \textbf{Good:}
    
    module adder(input [7:0] a, b, output [7:0] sum);  
        
        assign sum = a + b;  

    endmodule
    \end{minipage}
    \end{mdframed}
    \begin{mdframed}[backgroundcolor=lightred] 
    \begin{minipage}{\columnwidth}
    \textbf{Bad:}
    
    module adder(input logic [7:0] a, b, output logic [7:0] sum)  
    
        sum = a + b  

    endmodule
    \end{minipage}
    \end{mdframed}
    \vspace{5pt}

    \textit{The latter example includes SystemVerilog constructs in a .v file and has syntax errors.}
    \item \textbf{Diverse structural designs}, spanning arithmetic cores, crypto engines, DSP blocks, and more.
    \vspace{5pt}

    \begin{mdframed}[backgroundcolor=limegreen] 
    \begin{minipage}{\columnwidth}
    \textbf{Good:}
    
    A balanced mix of simple counters, medium-complexity crypto cores (e.g., AES), and multi-module DSP blocks.
    
    \end{minipage}
    \end{mdframed}

    \begin{mdframed}[backgroundcolor=lightred] 
    \begin{minipage}{\columnwidth}
    \textbf{Bad:}
    
    Lack of representative examples of the diverse module types, or contains unsynthesizable testbenches.
    
    \end{minipage}
    \end{mdframed}
    \vspace{5pt}

    \item \textbf{Reasonable formatting and indentation}, enabling the model to learn standardized yet diverse code structure.
    \vspace{5pt}

    \begin{mdframed}[backgroundcolor=limegreen] 
    \begin{minipage}{\columnwidth}
    \textbf{Good:}
    
    2- or 4-space indents, consistent naming (e.g., \texttt{clk}, \texttt{rst}, \texttt{data\_in}), and line spacing.
    
    \end{minipage}
    \end{mdframed} 

    \begin{mdframed}[backgroundcolor=lightred] 
    \begin{minipage}{\columnwidth}
    \textbf{Bad:}
    
    module X(input a,output b);assign b=~a;endmodule

    \end{minipage}
    \end{mdframed}
    \vspace{5pt}

    \item \textbf{Contextual metadata}, including I/O definitions, module names, and design comments.
    \vspace{5pt}

    \begin{mdframed}[backgroundcolor=limegreen] 
    \begin{minipage}{\columnwidth}
    \textbf{Good:}
    
    // 8-bit up-counter with asynchronous reset
    
    module counter(input clk, input rst, output reg [7:0] count);
    
    
        
        
    
    \end{minipage}
    \end{mdframed} 

    \begin{mdframed}[backgroundcolor=lightred] 
    \begin{minipage}{\columnwidth}
    \textbf{Bad:}
    
    No comments, or only boilerplate license text. I/O is ambiguous or undocumented.
    
    \end{minipage}
    \end{mdframed}
    \vspace{5pt}

    \item \textbf{Concise natural language descriptions}, which serve as prompts during instruction tuning.
    \vspace{5pt}

    \begin{mdframed}[backgroundcolor=limegreen] 
    \begin{minipage}{\columnwidth}
    \textbf{Good:}
    
    ``This module performs an 8-bit addition between two inputs and outputs the result.''
    
    \end{minipage}
    \end{mdframed} 

    \begin{mdframed}[backgroundcolor=lightred] 
    \begin{minipage}{\columnwidth}
    \textbf{Bad:}
    
    ``This file was created by XYZ. For internal use only.'' or overly verbose documentation.
    
    \end{minipage}
    \end{mdframed}

\end{itemize}
\vspace{-2pt}

    

In our work, we created a structured pipeline that reflects these findings. Data is obtained from trusted open-source repositories, cleaned using filtering and de-duplication tools, and validated via syntax and synthesis checks. We additionally extract metadata and descriptions for every module, allowing for prompt-response formatting compatible with instruction-tuned LLMs. The result is a dataset created specifically to support LLM training for RTL generation that is capable of generalizing across a range of module types and design idioms encountered in modern SoC development.

\subsection{Prior Work} \label{subsec:background_prior}

Research has shown LLMs are being applied to electronic design automation (EDA)~\cite{Huang_2021, ALSAQER2024, Yu_ML_CAD_2023}, design verification~\cite{autobench, zhang2025llm4dvusinglargelanguage}, hardware security~\cite{Saha2024, Saha2024_empowering, saha2025threatlensllmguidedthreatmodeling}, and HDL code generation~\cite{chipgpt, GPT4AIGChip}. Early foundational studies including \textit{ChipGPT}~\cite{chipgpt}, \textit{Chip-Chat}~\cite{chip-chat}, \textit{AutoChip}~\cite{AutoChip}, and \textit{GPT4AIGChip}~\cite{GPT4AIGChip} aimed to improve hardware design outputs produced by LLMs with prompt engineering. In recent research, there has been a trend towards domain-specific pretraining and fine-tuning, exemplified by works such as \textit{MG-Verilog}~\cite{mg}, \textit{RTLCoder}~\cite{rtlcoder}, \textit{Verigen}~\cite{verigen}, and \textit{VerilogEval}~\cite{verilogeval}. The latest efforts have sophisticated the research of fine-tuning and agentic approaches with large DBs and better frameworks. These include but are not limited to OriGen~\cite{origen}, BetterV~\cite{betterv}, AutoVCoder~\cite{autovcoder}, CodeV~\cite{codev}, CraftRTL~\cite{craftrtl}, and RTL++~\cite{rtlpp}.

\subsubsection{Prompt Engineering Approaches}
Various techniques enhance an LLM's RTL generation accuracy through prompt engineering. \textit{ChipGPT}~\cite{chipgpt} improves Verilog generation outputs from widely-used LLMs (e.g. ChatGPT) through a structured prompting scheme tailored for LLMs, enhancing code quality by incorporating hardware-specific context (e.g., input-output ports, module functions). It also cleaned the output codes, correcting and optimizing generated RTL, searching the design space and selecting optimal designs based on PPA metrics. While effective in elevating generation from general-purpose LLMs, ChipGPT relies solely on prompt engineering without fine-tuning, and lacks support for hierarchical or multi-module designs, limiting its applicability to complex systems. \textit{Chip-Chat}~\cite{chip-chat} utilizes a pre-trained LLM (like ChatGPT) without fine-tuning, and instead, depends on prompt engineering to steer code generation. The system incorporates a feedback mechanism in which user inputs enhance the outputs of the LLM. While it improves usability, \textit{Chip-Chat} does not incorporate fine-tuning or domain-specific training. As a result, the generated code often requires manual correction and may not consistently meet functional or structural correctness for complex or integrated designs. \textit{GPT4AIGChip}~\cite{GPT4AIGChip} utilizes tailored prompts to direct GPT-4, embedding hardware constraints and AI accelerator design concepts, such as data flow architectures and tiling strategies. It implements LLM-based parameter optimization (e.g., dataflow, bit-width) to enhance PPA metrics, while incorporating feedback from EDA tools to improve the generated designs. While this approach integrates architectural constraints and AI chip design knowledge, it lacks fine-tuning and depends entirely on handcrafted prompts. \textit{RTL++}~\cite{rtlpp} proposes a graph-enhanced generation method using control and dataflow graph embeddings to strengthen semantic understanding in Verilog synthesis. The model improves over token-based LLMs in capturing structural dependencies. However, it is primarily benchmarked on simplified designs, and the effectiveness of graph-guided generation in full-chip RTL is largely untested. 

\subsubsection{Fine-tuned Models} General-purpose models can be further enhanced by training them on domain-specific datasets. \textit{MG-Verilog}~\cite{mg} introduces a multi-grained dataset that addresses the limitations of existing corpora by providing Verilog code samples at varying levels of abstraction, which allows for more effective fine-tuning and in-context learning for LLMs. This dataset emphasizes the importance of diverse and well-annotated data in enhancing model performance. \textit{MG-Verilog} also tests this dataset by fine-tuning the CodeLLaMA-7B-Instruct model. However, \textit{MG-Verilog} has some limitations that may hinder its broader applicability. Despite containing over 11,000 samples, the dataset size remains relatively small for large-scale LLM training, potentially constraining generalization across different hardware design tasks. \textit{RTLCoder}~\cite{rtlcoder} is an RTL code generation framework that addresses the challenges of privacy concerns and the lack of high-performance open-source LLMs for Verilog generation by using a specialized training dataset and a customized model architecture. However, \textit{RTLCoder’s} scope is constrained by the model’s compact size. While it is efficient for deployment, the size limits its capacity to capture long-range dependencies or complex hierarchical design patterns often seen in industry-specific hardware designs. Moreover, even though the accompanying dataset supports syntax-based learning, it lacks coverage of advanced design descriptions such as finite-state machines (FSMs), pipelined architectures, or memory-mapped modules, which can reduce the model’s applicability. And, as with many recent LLM-based HDL tools, the framework primarily evaluates syntactic and token-level metrics. 

\textit{Verigen}~\cite{verigen} further contributes to the necessity of a proper dataset as it is a fine-tuned LLM explicitly optimized for Verilog code generation, achieving significant improvements in syntactic correctness and simulation-based pass rates over leading models. The authors fine-tune their model on a curated dataset sourced from GitHub repositories and educational materials, and evaluate across different Verilog generation tasks including module instantiation, conditional logic, and arithmetic design. Despite these advancements, however, \textit{Verigen} has limitations that constrain its generalizability. The training dataset, while carefully filtered, is primarily composed of academic-level and textbook-style designs, which may not reflect the structural and functional complexity found in industry-grade RTL. Additionally, it often fails to comprehend natural language within prompts, resulting in hallucination within simple codes on its smaller models. \textit{OriGen}~\cite{origen} introduces a RTL code generation framework, emphasizing self-reflection capabilities and an innovative code-to-code augmentation methodology. By using knowledge distillation from commercial LLMs like Claude3-Haiku, \textit{OriGen} refines open-source RTL code datasets to improve training quality. The framework employs a dual-model structure: Gen LoRA for initial code generation based on natural language instructions, and Fix LoRA for iterative error correction using compiler feedback. The drawback to this technique is that the augmentation strategies primarily focus on simple module patterns like adders and basic arithmetic units, limiting \textit{OriGen}'s applicability to more complex or hierarchical designs. 

\textit{BetterV}~\cite{betterv} builds upon compiler-feedback approaches by integrating generative discriminators and reward models tied to PPA. It attempts to improve generation quality with hardware metrics. However, its evaluation focuses on synthetic test cases, with limited exploration of simulation, assertion-based validation, or deployment scenarios. \textit{AutoVCoder}~\cite{autovcoder} introduces a two-stage fine-tuning pipeline combined with RAG to enhance both syntax and simulation-level correctness. It uses Verilog corpora along with code search to augment generation with applicable examples. This is effective in reducing syntax errors, but \textit{AutoVCoder} does not verify generated outputs through computer-aided design (CAD) tools, making its use in high-assurance applications uncertain. \textit{CodeV}~\cite{codev} focuses on improving dataset quality by generating description-code pairs for training and summarization. This enhances prompt construction and provides clearer supervision signals for LLMs during fine-tuning. However, the generated summaries are often generic, and the dataset lacks annotation for resource constraints or modular hierarchy. \textit{CraftRTL}~\cite{craftrtl} integrates graphical inputs such as waveforms and FSM diagrams to support multimodal Verilog learning. The model leverages error feedback during training to improve code correctness. While promising in its use of visual-verbal alignment, the training data is limited, and there is little evidence of success on complex control logic or multi-module designs. Its utility in traditional design flows remains experimental.

\subsubsection{Agentic Methods} Other approaches attempt to increase model performance through feedback mechanisms relying on integrating multiple LLM agents. \textit{AutoChip}~\cite{AutoChip} provides a framework where an LLM such as GPT-4o produces the initial Verilog code based on a natural language instruction, for instance, “create a 4-bit adder.” The code is then assessed for syntax errors using a Verilog compiler, like Verilator \cite{verilator}, and for functional accuracy through testbenches. Additionally, a feedback loop allows the LLM to refine the code iteratively, continuing until the code is correct or an iteration limit is reached. While \textit{AutoChip} demonstrates improved accuracy through automated refinement, it has limited applicability for large-scale hardware blocks. \textit{PromptV} or \textit{CoopetitiveV}~\cite{Coopetitivev} leveraged multi-agent prompting where multiple models handled different aspects of the code generation process. The goal is to create a collaborative yet competitive relationship between models to produce higher quality Verilog. With the increased overhead of many models, it requires considerations for scalability and deployment in resource-constrained environments. Also, high accuracy for complex hierarchical Verilog produced through integrating multiple LLM outputs is yet to be seen.

\subsubsection{Benchmarks}
\textit{VerilogEval}~\cite{verilogeval} is a multi-stage benchmarking framework to assess the capabilities of LLMs in Verilog code generation in a structured manner. This framework evaluates models based on syntax, functional simulation, and formal verification, allowing for consistent comparisons and helping identify areas for improvement. The benchmark is still limited by the small scale of its dataset. The benchmark primarily targets self-contained Verilog modules such as adders, multiplexers, and basic control logic, which do not reflect the complexity or hierarchical depth of real-world hardware designs. As a result, models evaluated using \textit{VerilogEval} may show strong results on these micro-benchmarks while struggling with larger, pipelined modules common in industrial RTL. Additionally, the dataset lacks diversity in coding style, parameterization, and integration with IP cores or testbenches, which are necessary for assessing generalization. These constraints limit the benchmark’s ability to capture practical challenges during synthesis, integration, or in the bitstream. Despite the limitations, \textit{VerilogEval} is a great start for DB creation because it puts criteria in place for proper evaluation of LLM Verilog code generation. Our goal is to extend the current literature with an automated framework of curating, preprocessing, and storing a large high-quality Verilog dataset for effective GenAI fine-tuning. \textit{RTLLM}~\cite{lu2024rtllm,rtllm} is a benchmark and evaluation framework that currently includes 50 hand-crafted designs that include hand-made RTL designs, their testbenches, and natural language descriptions. The evaluation framework facilitates syntax checks, functionality checks via simulations, and design quality checks (PPA) post-synthesis. These 50 designs, although high-quality, may not adequately test the full complexity and diversity of real world designs.

The field of LLM-driven Verilog generation has rapidly evolved from early prompt-engineering-based approaches toward more sophisticated frameworks that incorporate fine-tuning, compiler feedback, and benchmarking. Despite these advances, significant challenges remain, including small dataset sizes, limited support for complex hierarchical designs, and inconsistent verification of generated codes. The latest tools attempt to improve these drawbacks by exploring multimodal inputs, reward-based training, and graph-enhanced modeling. However, they still fall short in terms of real-world applicability, scalability, and full-system integration. This highlights the need for a large-scale dataset curation framework that enables proper domain-specific fine-tuning or agentic approaches for more consistent efforts in RTL code generation.

\section{VerilogDB}
\label{section:verilogdb}

To allow scalable training and evaluation of LLMs in Verilog code, we developed \textit{VerilogDB}, an end-to-end system that automates the collection, sanitization, annotation and structuring of Verilog RTL designs. This framework is designed to collect diverse sources of Verilog code, apply filtering and synthesis-aware validation, and output standardized metadata and instruction-friendly formats. The pipeline spans from raw repository scraping and PDF extraction to structured storage in a DB that can be queried. The following subsections detail each major component of \textit{VerilogDB}, including our data collection strategy in Section~\ref{subsec:data_collection}, preprocessing filters in Section~\ref{subection:preprocessing}, and DB architecture in Section~\ref{subsec:database_arch}.

\subsection{Data Collection}\label{subsec:data_collection}

\begin{figure}[htbp]
\centering
\fbox{\includegraphics[width=0.9\linewidth]{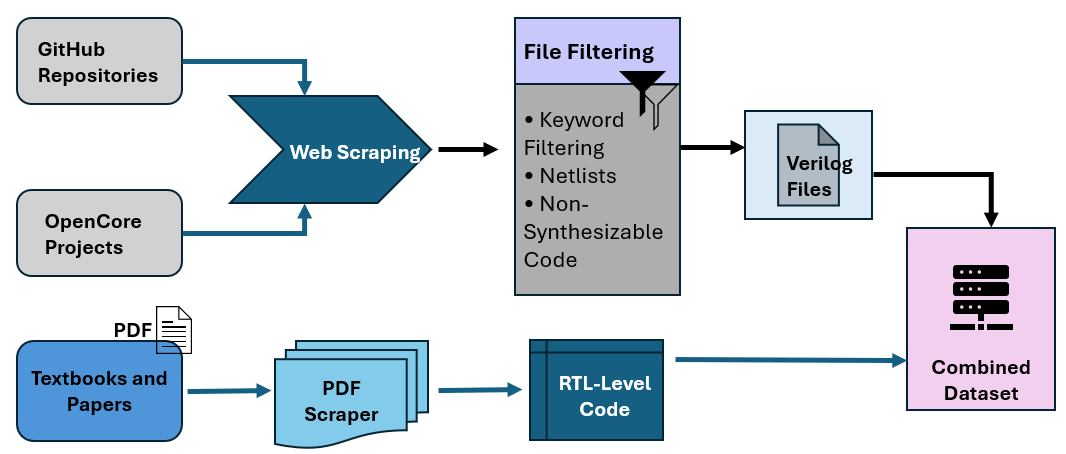}}
\vspace{-8pt}
\caption{Overview of the Verilog RTL data collection process from GitHub, OpenCores, and textbooks. Each source passes through a source-specific pipeline to extract syntactically valid and synthesizable RTL code.}
\vspace{6pt}
\label{fig:data-collection-flow}
\end{figure}

A critical component in constructing a large, diverse, and high-quality dataset is curating the data samples adequately to meet the goal of consistent LLM generation of Verilog. To meet this requirement, we developed a data collection framework targeting publicly available Verilog repositories. Our sources included general-purpose hosting platforms (e.g., GitHub), hardware-specific archives (e.g., OpenCores), and educational repositories hosted on academic domains including and not limited to digital design textbooks. The overarching goal was to capture real-world design diversity while guaranteeing structural clarity before the preprocessing stage explained in Section~\ref{subection:preprocessing}.

\subsubsection{GitHub Scraping}

We queried GitHub through its REST API to identify repositories relevant to hardware design. Using keyword filters such as \texttt{verilog}, \texttt{rtl}, \texttt{fpga}, \texttt{hdl}, and \texttt{asic}, we retrieved a corpus of potential repositories from multiple sources and users. Each repository was then cloned, and only those containing Verilog source files (\texttt{.v}) were retained. To enforce relevance and quality, we implemented a branch-level and folder-level exclusion mechanism where branch names and directory paths containing generic or infrastructure-focused keywords were discarded to avoid including test harnesses, simulation environments, or system-level wrappers. 
Within each selected repository, Verilog files were extracted along with accompanying documentation (e.g., \texttt{README}, license files, and comments). The data collected from GitHub amounted to more than $\sim$30 GB of raw data that span a wide range of application domains, code styles, and project sizes, from single-module educational designs to hierarchical industrial IPs. After filtering, $\sim$2 GB of Verilog data with around 72,600 Verilog modules are prepared and passed through the preprocessing pipeline.

\subsubsection{OpenCores and Academic Sources}

OpenCores~\cite{opencores} is one of the longest-standing archives of open-source HDL IPs and serves as a prominent source of synthesizable Verilog modules. We examined project pages across categories such as arithmetic cores, cryptographic primitives, and DSP blocks. Each project was downloaded and filtered through our standard exclusion criteria, targeting the elimination of:
\begin{itemize}
    \item Files labeled as netlists or synthesis output (e.g., \texttt{\_netlist.v}, \texttt{\_synth.v})
    \item Testbenches or simulation artifacts (e.g., files containing \texttt{\$display}, \texttt{initial}, \texttt{waveform.vcd}, or \texttt{dumpfile})
    \item Projects using unsupported language extensions (e.g., SystemVerilog in plain \texttt{.v} files)
\end{itemize}

Approximately 165 MB of raw Verilog was collected from OpenCores, which, after filtering, yielded 874 clean RTL modules suitable for synthesis and further annotation.

We further complemented our corpus by mining Verilog samples from academic course websites. These sources included assignment repositories, lab guides, and instructor repositories from top-ranked universities offering courses in digital logic or computer architecture. Although smaller in volume, these files are often well-commented and follow canonical design practices, making them ideal training examples. We extracted $\sim$500KB of data samples from these sources.

\subsubsection{Organization, Challenges, and Raw Statistics}

All files were grouped by source and repository, preserving original folder hierarchies. This is significant for the later synthesis testing (discussed in Section \ref{subsec:synth}). We maintained metadata for each project, including:
\begin{itemize}
    \item Source of origin (e.g., GitHub, OpenCores, Academic)
    \item File type distributions (e.g., Verilog, testbench, config, README)
    \item Repository-level notes such as license, readme presence, and author information
\end{itemize}

One of the primary challenges encountered during this phase was the high variability in Verilog code quality and structure. Some repositories used informal or undocumented coding styles, lacked top-level modules, or contained extensive non-Verilog assets (e.g., C test drivers, synthesis scripts, IPXACT metadata). We used a permissive initial collection strategy to avoid early bias, deferring aggressive filtering to the preprocessing stage (Section~\ref{subection:preprocessing}).

In total, the aggregated raw dataset consists of:
\begin{itemize}
    \item \textbf{GitHub}: $\sim$30 GB pre-filtered across thousands of repositories and 2 GB filtered
    \item \textbf{OpenCores}: $\sim$165 MB of raw Verilog IP cores
    \item \textbf{Academic sources}: $\sim$500KB manually verified course-level RTL projects from top digital design curricula
\end{itemize}

This collection forms the foundation to our structured preprocessing pipeline, designed to provide high fidelity and LLM compatibility for instruction tuning and Verilog generation.

\subsection{Data Preprocessing}
\label{subection:preprocessing}

\begin{figure*}[t]
\centering
\fbox{\includegraphics[width=0.95\columnwidth]{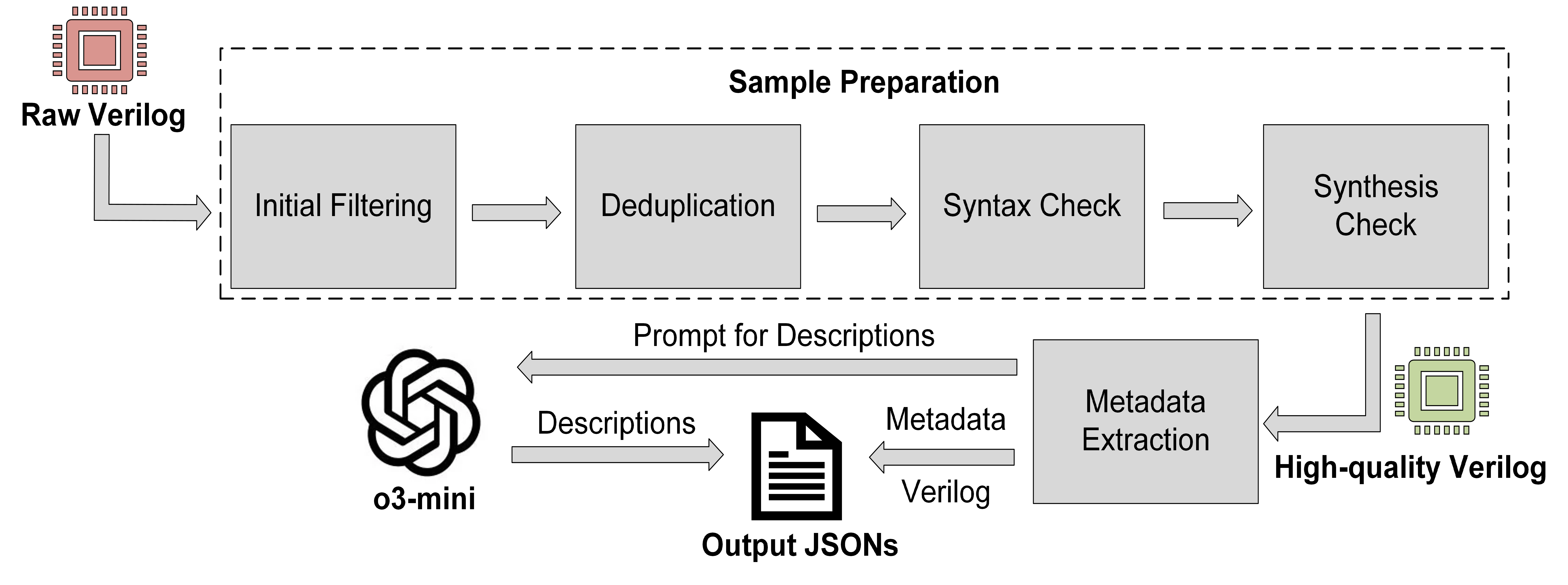}}
\vspace{-8pt}
\caption{Overview of the VerilogDB Framework Data Preprocessing Flow.}
\vspace{6pt}
\label{fig:preprocessing-flow}
\end{figure*}

The preprocessing pipeline addresses challenges in Verilog codebases, such as duplicate files, syntax errors, and synthesis issues, while extracting critical module metadata (e.g., module names, ports, comments, and token counts) and formatting it into a JSON-based structure optimized for LLM consumption. The goal of this flow is to assure the quality of data samples for effective LLM fine tuning. Figure \ref{fig:preprocessing-flow} illustrates the stages involved in the preprocessing pipeline of \textit{VerilogDB}. The workflow consists of five primary steps: (1) initial file filtering, (2) duplicate removal, (3) syntax checking, (4) logic synthesis, and (5) metadata extraction. Each stage is designed to maintain data quality and compatibility with downstream LLM applications, such as code generation, hardware design analysis, or automated documentation.

\subsubsection{Initial File Filtering}
As mentioned in Section~\ref{subsec:data_collection}, a preliminary filtering removes files which do not fit the use case or to reduce processing overhead. If the files or project folders contain *\_netlist.v, *\_gate.v, *\_mapped.v, or *\_synth.v, they are removed from the initial sample as the dataset should only include valid Verilog modules useful for LLM fine tuning for RTL generation. Likewise, if the data contained sim, simulate, waveform, \_test, they were also removed as we desire solely .v files which are synthesizable module files. The script excludes testbenches by filtering files containing "tb", "TB", or similar testbench nomenclature in their names and modules, as these typically include non-synthesizable constructs like delays or file I/O. 

\subsubsection{Deduplication}
The second stage of the preprocessing pipeline focuses on eliminating duplicate Verilog files to reduce redundancy and improve processing efficiency in the later stages. Deduplication improves the quality of a dataset, reduces bias, and improves the performance of the model when fine-tuned or trained. The fine-tuning dataset must be generalizable and representative, avoiding concerns like overfitting and memorization. Duplicate Verilog files can arise in large HDL repositories due to version control artifacts, automated design tools, or copied modules. A custom Python deduplication script was built and deployed, utilizing MD5 hashing~\cite{md5} for accurate detection of identical files based on their content, regardless of file names or locations. Algorithm~\ref{alg:dupe} shows the \textit{VerilogDB} deduplication process. MD5 was selected for its speed, ease of implementation, and efficiency. As seen in lines \ref{line:md5_hash}-\ref{line:md5_hash_2}, the script first iterates through the designated project directory, computing a hash for each .v file and grouping files by their hash values. For each group of duplicates, the script retains the file with the shortest path (to prioritize original sources) and removes the others from subsequent processing as shown in line \ref{line:keep_file_dupe}. The deduplication stage produces a unique set of Verilog files, minimizing computational overhead in subsequent stages and reducing overfitting in LLM training due to redundant data~\cite{Sheik_2024, Lee_dedup_2022}. 

\begin{algorithm}
  \KwRequire{$input\_dir$}
  \KwResult{$output\_dir$}

  \tcp{Initialize dictionary for storage of file paths by hash}
  $\Var{hash\_to\_files} \gets \Var{empty dictionary}$\;
  \tcp{Step 1: Collect all .v files and compute their hashes}
 
  \For{$\Var{file}$ in $\Var{recursive traversal}$ of $input\_dir$}{%
    \label{line:md5_hash}
    \If{is\_verilog\_file($\Var{file}$)}{%
        $\Var{file\_path} \gets \Var{full path to file}$\;
        $\Var{file\_hash} \gets \Var{\textit{ComputeMD5Hash}(file\_path)}$\;
        \If{file\_hash exists in hash\_to\_files}{
        Append file\_path to hash\_to\_files[file\_hash]
        } 
        \Else{
        $\Var{hash\_to\_files[file\_hash]} \gets \Var{new list containing file\_path}$\;
    
        }
    }
    }\label{line:md5_hash_2}
    \tcp{Step 2: Copy one file per unique hash to output directory}

    \For{$\Var{file\_hash}$,$\Var{file\_paths}$ in $\Var{hash\_to\_files}$}{%
    $\Var{kept\_file} \gets \Var{file\_paths[0]}$\; \label{line:keep_file_dupe}
    $\Var{rel\_path} \gets \Var{relative path of kept\_file with respect to input\_dir}$\;
    $\Var{output\_path} \gets \Var{\textit{join}(output\_dir, rel\_path)}$\;

    \tcp{Create necessary directories}
    Create directory structure for output\_path if not exists
    
    \tcp{Copy file to output directory}
    Copy kept\_file to output\_path with metadata preservation

    }
 \caption{Duplicate Removal Algorithm}
  \label{alg:dupe}
\end{algorithm}

\subsubsection{Syntax Check}
\label{subsec:synth}
Following deduplication, the pipeline performs syntax checking to verify that Verilog files adhere to the HDL's grammatical rules. Syntax errors, such as missing semicolons, improper module declarations, and missing module names in instantiations, can prevent successful compilation and synthesis, rendering files unsuitable for metadata extraction and for the dataset as a whole. A custom Python script utilizes the Icarus Verilog (iVerilog) compiler~\cite{iverilog}, a widely-used open-source tool, to verify each code's syntax. It automated the syntax checking process, allowing for a quicker preprocessing flow than manual inspection of each data sample. Algorithm~\ref{alg:syntax} shows the \textit{VerilogDB} process for the syntax checking per each Verilog file. The script processes each .v file in the deduplicated directory, invoking iVerilog in subprocesses, seen in lines \ref{alg:icarusv}-\ref{alg:icarusv2}, to check syntax for each file without generating output binaries. The goal of this stage is syntax check and not elaboration, which will be verified in the later synthesizability check stage. Responses from iVerilog help indicate the error/warning types. Moreover, the script properly isolates syntax errors from elaboration errors, creating a list of the files which pass syntax and ignores the elaboration errors as we do not check synthesis yet to increase processing efficiency. This can be seen in lines \ref{alg:syntax_errors}-\ref{alg:syntax_errors2}. These files that pass the syntax check are copied to an output directory, while errors are logged with details of the failure (e.g., line numbers and error messages).

\begin{algorithm}
  \KwRequire{$file\_path$}
  \KwResult{$success$}

  \tcp{Verify tool and file exist}
  Leave algorithm if tools and file does not exist.

  \tcp{Run Icarus Verilog with syntax check options}
  $\Var{cmd} \gets \Var{[IVERILOG\_PATH, "-o", "nul", "-Wall", file\_path]}$\; \label{alg:icarusv}
  $\Var{result} \gets \Var{execute cmd, capture stdout and stderr as text}$\; \label{alg:icarusv2}
  $\Var{stderr\_output} \gets \Var{strip result.stderr}$\;

  \tcp{Check for clean pass}
  \If{result.returncode = 0}{
  
        \If{stderr\_output is not empty}{
                Log "Warnings detected: stderr\_output"
                
                  \KwRet{True, "Syntax check passed (with warnings: stderr\_output)"}
        }
        \Else{
                \KwRet{True, "Syntax check passed"}
        }
        
   }

   \tcp{Analyze errors}
   $\Var{error\_lines} \gets \Var{split stderr\_output into lines}$\;
   $\Var{has\_syntax\_error} \gets \Var{False}$\;
   $\Var{elaboration\_errors} \gets \Var{empty list}$\;
   \tcp{Define error patterns}
   $\Var{syntax\_error\_pattern} \gets \Var{regex matching syntax error outputs from iverilog}$\; \label{alg:syntax_errors}
   
   $\Var{elaboration\_error\_pattern} \gets \Var{regex matching elaboration error outputs from iverilog}$\;
                 
   \For{$\Var{line}$ in $\Var{error\_lines}$}{%

        \If{syntax\_error\_pattern matches line}{
         $\Var{has\_syntax\_error} \gets \Var{True}$\;
         }
        \ElseIf{elaboration\_error\_pattern matches line or "warning: macro" in line}{
                Append line to elaboration\_errors
        }
    }
        
    \tcp{Determine outcome}
        \If{has\_syntax\_error and elaboration\_errors is empty}{
            $\Var{error\_message} \gets \Var{stderr\_output or "Unknown syntax error"}$\;
            \KwRet{False, "Syntax error: error\_message"}
        }

        \ElseIf{elaboration\_errors is not empty}{
            $\Var{error\_message} \gets \Var{join elaboration\_errors with newlines}$\;
            \KwRet{True, "Syntax OK, elaboration issues: error\_message"}
        }
        \Else{
            $\Var{error\_message} \gets \Var{stderr\_output or "Unknown error"}$\;
            \KwRet{False, "Check failed: error\_message"}
        } \label{alg:syntax_errors2}
    
  \caption{Syntax Check Algorithm per .v File}
  \label{alg:syntax}
\end{algorithm}

\subsubsection{Synthesizability Check}
After cleaning the data samples from netlists, testbenches, duplicates, and syntactically-incorrect code, the third stage verifies the synthesizability of the Verilog files. Synthesis checking is critical for LLM applications targeting hardware design, as non-synthesizable codes present in the dataset will skew LLMs' fine-tuning or training away from generating code that can be used for a physical device. We developed a script that employs Yosys~  \cite{yosys}, an open-source synthesis tool, to perform synthesis checks on each subproject. Algorithm~\ref{alg:synthesis} shows the \textit{VerilogDB} process for checking synthesizability per project. The script operates at the repository/project-level with the goal of testing multiple files at once for synthesizability, as Yosys facilitates batch processing projects for logic synthesis which is the standard for these tools.

Firstly, a top-level module is identified from the subproject, finding the hierarchy (lines \ref{alg:gettop}-\ref{alg:gettop2}). The script does this by first checking the files with "top" in the module or file name and then iteratively checking for which file in a project will synthesize as the top-level. This iteration is necessary as many times top modules do not have "top" in the file name. As seen in lines \ref{alg:yosys_flow}-\ref{alg:yosys_flow2}, each syntax-checked file within the scope of a subproject is batch processed using Yosys's synth command, which attempts to map the Verilog code to a generic gate-level netlist. Files in projects that are successfully synthesized are passed to the metadata extraction stage and failures are logged with diagnostic information. We also note that there were cases where projects had files with the same module names which the script handles by checking synthesizability of all possible scenarios of modules with that same name. For example, a project may have two modules named $counter$ but one counts to 256 than the other 512. The script checks synthesizability for each case where either counter is instantiated in the design.

\begin{algorithm}
  \KwRequire{$files\_in\_proj$}
  \KwResult{$failure, output$}

  \tcp{Setup and tool/file check}
   Leave algorithm if Yosys and file does not exist.
   
  \tcp{Identify unique modules}
   $\Var{unique\_files} \gets GetUniqueModules\Var{(files\_in\_proj)}$\; \label{alg:gettop}
   $\Var{top\_module} \gets DetermineTopModule\Var{(unique\_files)}$\; \label{alg:gettop2}

  \tcp{Build Yosys script}
   $\Var{script} \gets \Var{empty string}$\; \label{alg:yosys_flow}

   \For{$\Var{file}$ in $\Var{unqiue\_files}$}{%
        Append "read\_verilog \textbackslash{}"file\textbackslash{}"" to script
    }
    Append "hierarchy -check -top top\_module" to script
    
    Append "synth -top top\_module" to script

   \tcp{Run Yosys}
   $\Var{cmd} \gets \Var{[YOSYS\_PATH, "-q", "-p", script]}$\;
   $\Var{result} \gets \Var{execute cmd, capture stdout and stderr as text}$\;
   $\Var{output} \gets \Var{\textit{strip}(result.stdout) + \textit{newline}() + \textit{strip}(result.stderr)}$\; \label{alg:yosys_flow2}

   \tcp{Analyze Errors}
   $\Var{errors} \gets \Var{matches of regex pattern for non-tolerable Yosys error/warning outputs}$\;
   \If{errors is empty}{
         \KwRet{False, output}
         \tcp{No Errors, Synthesis Passed}
    }
    \Else{
         \KwRet{True, output}
         \tcp{Errors Present, Synthesis Failed}
    }
  \caption{Synthesis Check Algorithm per Project}
  \label{alg:synthesis}
\end{algorithm}
\vspace{-18pt}

\subsubsection{Metadata Extraction}


\begin{figure}[h]
\centering
\begin{lstlisting}[language=json, frame=single]
{
  "module_name": "dec",
  "ports": [
    {
        "name": "I",
        "direction": "input",
        "bit_width": 2
    },
    {
        "name": "v",
        "direction": "input",
        "bit_width": 1
    },
    {
        "name": "y",
        "direction": "output",
        "bit_width": 4
    }
  ],
  "comments": [],
  "verilog_code": "module dec (\n input [1:0] I,\n input v,\n output reg [3:0] y\n);\n\n always@(I)\n begin\n case({I,v})\n 3'b001: y = 4'b0001;\n 3'b011: y = 4'b0010;\n 3'b101: y = 4'b0100;\n 3'b111: y = 4'b1000;\n default: y=4'b0000;\n endcase\n end\nendmodule",
  "token_count": 35,
  "description": "This module decodes a 2-bit input and control signal into one of four 4-bit outputs."
}
\end{lstlisting}
\vspace{-8pt}
\caption{Data sample JSON for decoder containing metadata and generated description.}
\label{lst:dec}
\end{figure}

The metadata extraction stage for sample preparation, implemented in a custom Python script, processes the synthesizable Verilog files, capturing specific metadata from each module including the module name, port names and widths, comments, estimated LLM token count, and a high-level natural language description of the code. The primary challenge of this stage is parsing diverse Verilog syntax to capture essential module information, including module names, port declarations, comments, parameters, and estimated token counts. The script employs regular expressions (regex) to robustly handle Verilog constructs, ensuring compatibility with different coding styles and edge cases.
\begin{enumerate}
    \item \textit{Regex-based Parsing} - Firstly, modules are identified in the .v file. A regex pattern $MODULE\_RE$ matches module declarations while ignoring commented-out modules (e.g., //module ...) using negative lookbehinds ((?<!//.*)). This ensures only active modules are processed, addressing issues where commented code may be mistakenly parsed. Also, in the module declaration, whitespace is taken into account as the definitions may differ in the use of whitespace. We did not normalize the whitespace in preprocessing because we wish for the LLM to understand the variety of spacing styles and indentation practices. It also identifies cases where a Verilog file has multiple modules. Next, a regex pattern $PORT\_BLOCK\_RE$ captures input, output, and inout ports, supporting both ANSI and Non-ANSI styles and the various port definition formats (e.g., comma-separated port definitions, bit-widths with or without spaces, etc) and keywords (e.g., wire, reg, signed). Comments are also extracted with a regex pattern $COMMENT\_RE$ identifying single-line (// ...) and multi-line (/* ... */) comments. In the case where the module definition has parameter declarations (e.g., parameter WIDTH = 8;) for bit-width calculations, these are identified with $PARAMETER\_RE$ regex pattern.
    \item \textit{Token Estimation and Description Generation} - Aside from a conventional Verilog parser/tokenizer which are syntax-aware and may not be accurate to how LLMs interpret tokens, we developed a custom tokenizer to estimate the number of tokens as LLMs would (e.g. handling identifier names, whitespace, and punctuation). The arduous process of annotating module descriptions was also automated to greatly scale the DB using OpenAI's o3-mini to generate for every module, a corresponding singular sentence description with a maximum of 40 words. A short description was utilized as to provide the LLM with a concise and efficient description. We prompted o3-mini with the following:

\vspace{3pt}

\begin{itemize}
    \item \textbf{Description Prompt to o3-mini:} Describe what the following Verilog code does in 40 words or less, ending with a period: \textbackslash{}n\textbackslash{}nf"{verilog\_code}\textbackslash{}n\textbackslash{}n" Focus on the module's core function.
\end{itemize}
\vspace{3pt}

\item \textit{Output Structure} - The extracted metadata is formatted as JSON files with one file per module (e.g., AES\_Core.json). Figure~\ref{lst:dec} shows an example outputted JSON for a D flip flop Verilog file d\_ff.v.  Each JSON includes:
\begin{itemize}
    \item \textit{module name}: The extracted name of the module.
    \item \textit{ports}: A list of dictionaries with name, direction, and bit\_width (e.g., [{"name": "din", "direction": "input", "bit\_width": 128}, ...]).
    \item \textit{comments}: A list of cleaned comments.
    \item \textit{Verilog code}: The full module code, preserving original formatting and comments.
    \item \textit{token count}: An estimate of code complexity based on whitespace and separator splitting. Useful for LLM fine-tuning.
    \item \textit{description}: The module's generated description from o3-mini.
\end{itemize}
\end{enumerate}

\subsection{Database Architecture} \label{subsec:database_arch}
With the sheer volume of needed Verilog data, a DB hosted on dedicated hardware with sufficient storage is a key part of our framework.  

\subsubsection{Database Selection} \label{subsubsec: database_selection}
PostgreSQL~\cite{postgresql} is our chosen platform for hosting the \textit{VerilogDB} dataset due to the following reasons:

\begin{enumerate}

\item \textit{Structured and Semi-Structured Data Support} - It handles exceptionally structured data (e.g., where module metadata is stored in tables) and semi-structured data (e.g., where module metadata is stored directly as JSONs). The \textit{VerilogDB} preprocessing flow generates JSON metadata, which PostgreSQL can efficiently store using its native JSONB data type.

\item \textit{Verilog Module Querying Capabilities} - It supports complex SQL queries, including joins, aggregations, and window functions, which are essential for analyzing \textit{VerilogDB}'s dataset (e.g., grouping by module complexity) as its full-text search and JSONB querying is an efficient retrieval of prompt-response pair instructions for LLM fine-tuning.

\item \textit{Scalability and Performance} - As it can scale vertically, PostgreSQL benefits \textit{VerilogDB} with its 20,392 module entries as the system handles large datasets efficiently with features like parallel query execution, indexing (e.g., B-tree, GIN for JSONB), and partitioning.

\item \textit{Data Integrity and Reliability} - Through constraints (e.g., primary keys, foreign keys, unique constraints), PostgreSQL enforces strong data integrity, guaranteeing that module metadata is consistent and reliable. Its ACID (Atomicity, Consistency, Isolation, Durability) compliance assists data management to preserve dataset quality during LLM fine-tuning.
\end{enumerate}

\subsubsection{DB Structure and Sample Insertion}



\begin{figure}[t]
\begin{lstlisting}[style=sqlstyle]
CREATE TABLE verilog_modules (
    id SERIAL PRIMARY KEY,
    module_name TEXT NOT NULL,
    verilog_code TEXT NOT NULL,
    description TEXT NOT NULL,
    comments TEXT,
    token_count INTEGER,
    extracted_at TIMESTAMP DEFAULT CURRENT_TIMESTAMP
);

-- Table to store ports for each module
CREATE TABLE module_ports (
    id SERIAL PRIMARY KEY,
    module_id INTEGER REFERENCES verilog_modules(id) ON DELETE CASCADE,
    port_name TEXT NOT NULL,
    port_type TEXT CHECK (port_type IN ('input', 'output', 'inout')) NOT NULL,
    port_width INTEGER,
    CONSTRAINT unique_port UNIQUE (module_id, port_name)
);
\end{lstlisting}
\vspace{-10pt}
\caption{SQL schema for PostgreSQL structured DB Implementation}
\label{lst:schema}
\end{figure}

As mentioned in Section~\ref{subsubsec: database_selection}, the PostgreSQL implementation can support both structured and semi-structured data storage. Figure~\ref{lst:schema} shows the SQL schema for the structured data storage implementation using PostgreSQL. Primary keys for the module IDs and port IDs provide unique identifiers that are B-tree indexed~\cite{bayer1972organization}, allowing for efficient lookups and joins that are critical for the retrieval of specific modules or the linkage of ports to modules. Data integrity is enforced where certain metadata (e.g., module names, descriptions, and port names) is required to be present. In the \texttt{module\_ports} table, each port in a module must be unique, further enforcing the Verilog structural rules. Comments from the Verilog codes are left as optional in the DB, as designers may or may not annotate their modules with comments. 

Automated scripts were developed to insert the JSON contents into the PostgreSQL DB tables in a streamlined fashion. Due to the selection of PostgreSQL, a variety of significant validation logic is inherent for proper metadata construction before DB insertion. We use these innate integrity benefits of the PostgreSQL DB to verify the metadata extraction and deny any insertion of improper entries. As ports are enforced to not be null before the module's insertion into the DB, any module where its ports or module definition were not captured by the regex patterns is denied entry into the DB. Any attempt to insert identical data to an entry in the DB is denied due to an added unique constraint on the \texttt{verilog\_code} column. Certain exceptions are made for modules that are meant to have no ports. For example, modules that are standard wrappers for fills or decoupling capacitors may contain no ports. 

\subsection{Sample Preparation for Instruction-Tuning}
JSON data for fine-tuning is designed to support LLM tasks such as code completion, design optimization, and automated documentation. The data can also aid in our purpose of RTL code generation. As a key final step for fine-tuning, we formatted the extracted metadata for LLM consumption via a custom Python script, which processes DB samples, providing compatibility with LLM training pipelines by formatting them into proper instructions~\cite{ANISUZZAMAN2025100184,parthasarathy2024ultimateguidefinetuningllms}. LLM instruction-tuning requires the data to be formatted in prompt-response pairs. For example, to instruction-tune an LLM that can effectively translate English to Spanish, the prompt would be to translate a phrase with the associated response as the translation. For the use case of Verilog code generation, the prompt would ask for a design following a set of specifications with the response being the code itself. The construction of the prompt component lies in the metadata. For each module prompt, we first place a consistent system prompt for the LLM~\cite{sahoo2025systematicsurveypromptengineering, schulhoff2025promptreportsystematicsurvey}, then prompt the LLM to generate a Verilog module named \texttt{module\_name}, including the ports and description of the module. Below is an example prompt and response for an adder circuit:

\vspace{6pt}
\begin{mdframed}[backgroundcolor=minipagecolor] 
\begin{minipage}{\columnwidth}
\vspace{2pt}
\textbf{Example Prompt $\mathbb{P}_1$:} You are a highly experienced RTL code designer skilled at designing concise, syntactically correct, and synthesizable Verilog code that functions. 

Generate Verilog code for a module named ADD with the following ports and description:

input [15:0] in1, input [15:0] in2, output [15:0] out

This module performs combinational addition, computing the sum of two 16-bit inputs to produce a 16-bit output.
\end{minipage}

\end{mdframed}
\begin{mdframed}[backgroundcolor=lightgray] 
\begin{minipage}{\columnwidth}
\vspace{2pt}
\textbf{Example Response $\mathbb{R}_1$:}

module ADD  (out, in1, in2);

input [15:0] in1, in2;

output reg [15:0] out;

always @(*)

out = in1 + in2; 

endmodule
\end{minipage}
\end{mdframed}
\vspace{6pt}

With the stored estimated token count value for each data element, data samples for specific context window sizes can be pulled. For example, for Mistral-7B~\cite{mistral7b} and Code Llama-7B~\cite{roziere2024code}, the context windows are 8,192 and 4,096 tokens respectively. This implies that Verilog code samples should ideally be below those token counts, as an LLM interpreting a truncated input code leads to a loss of information/context and thus degrades performance in many scenarios~\cite{wu-etal-2024-extending, shi2025explainingcontextlengthscaling}. 
\textit{VerilogDB}'s distribution of code based on token count is discussed and shown in Section~\ref{section:evaluation}.

\section{Analysis} \label{section:evaluation} 

We passed our $\sim$30 GB raw data through our preprocessing pipeline (see Figure \ref{fig:preprocessing-flow}), where the sample cleaning metrics per stage can be seen in Table \ref{tab:preprocessing_stages}. We note an effective processing flow yielding 751 MB of total output data size and 20,392 Verilog modules prepared for further LLM fine-tuning or other downstream tasks. In addition to an analysis on the preprocessing efficacy, we also assess \textit{VerilogDB's} data diversity along four key aspects: 1) Functional categories, 2) Module Complexity, and 3) Annotation and Comments. We also provide a discussion on the benefits and challenges of \textit{VerilogDB}.

\subsection{Preprocessing Evaluation}


\begin{table}[ht]
\centering
\caption{Verilog Preprocessing Pipeline Stages and Filtering Statistics}
\label{tab:preprocessing_stages}
\vspace{-10pt}
\begin{tabularx}{\linewidth}{|l|X|r|r|r|X|}
\hline
\textbf{Stage} & \textbf{Description} & \textbf{Input (MB)} & \textbf{Output (MB)} & \textbf{\% Retained} & \textbf{Common Rejection Reasons} \\
\hline
Deduplication & Removes textually identical modules & 2000 & 1750 & 87.53\% & Identical files \\
\hline
Syntax Check & Drops modules with syntactic errors during parsing & 1750 & 1104 & 63.07\% & Lexical issues, malformed declarations, incorrect use of blocking assignments \\
\hline
Synthesis Check & Passes only synthesizable modules & 1104 & 767 & 69.49\% & Unsupported logic, invalid constructs, use of delays \\
\hline
DB Validation & Ensures compatibility with internal database schema & 767 & 751 & 99.01\% & Schema mismatch, unresolved symbols, incorrectly parsed metadata \\
\hline
\end{tabularx}
\end{table}

After 2 GB, about 72,600 modules, passed initial filtering, this subset was passed through our remaining preprocessing pipeline. Approximately 12.47\% of the 72,600 modules were found to be exact duplicates and were removed, reducing the size of the dataset without losing unique design information. In our experiments, 63.07\% of deduplicated modules passed syntax checking, with failures primarily due to incomplete modules or non-standard Verilog constructs, which were excluded from further processing to maintain data quality. In our dataset, after removing testbenches and files with non-synthesizable constructs, approximately 69.49\% of syntax-checked modules passed synthesis. Common failures were attributed to unsupported constructs or missing dependencies, which were flagged for manual inspection. Metadata extraction was done successfully for over 99.01\% synthesized modules, with failures mainly due to improper port declarations and unidentified modules. These were caught and prevented for insertion by the DB validation logic. In our experiments, the preprocessing pipeline outputted a dataset of 751 MB of high-quality Verilog data, producing 20,392 unique, synthesizable modules with complete metadata, demonstrating diversity and scalability.

\subsection{Dataset Distribution and Diversity}

After deduplication and filtering, we examined the distribution and composition of the resulting Verilog dataset along functional, structural, and annotation-based axes. Our goal was to guarantee the retained modules reflect a wide variety of RTL design styles and application domains and span an array of sizes/complexity.

\subsubsection{Functional Categories.} 
\begin{figure}[h]
\centering
\fbox{\includegraphics[width=0.7\linewidth]{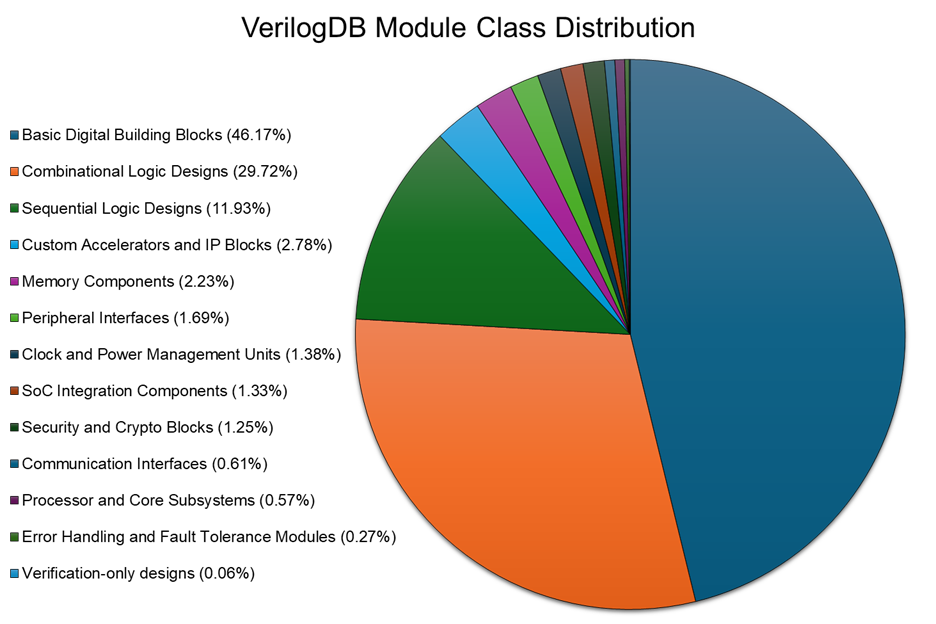}}
\caption{Class Distribution of \textit{VerilogDB} modules.}
\label{fig:classes}
\end{figure}

We observed a broad representation of modules across key RTL design classes. These include arithmetic blocks (e.g., adders, multipliers, ALUs), cryptographic components (e.g., AES rounds, modular multipliers), DSP modules (e.g., FIR filters, MAC units), control logic (e.g., counters, FSMs, encoders), and memory interfaces (e.g., register files, FIFOs, RAM controllers). This functional diversity is important for exposing LLMs to different types of datapaths, control patterns, and signal-level interactions typical in modern SoC designs. In order to adequately analyze the functional makeup of the dataset, we passed all \textit{VerilogDB's} modules to o3-mini to classify the modules into 13 different classes based on module type/functionality. Figure \ref{fig:classes} shows the distribution by functional class for the modules in \textit{VerilogDB}. The class numbers, names, and example modules are listed in Table \ref{tab:class_table} in the Appendix. Most modules in \textit{VerilogDB} fall under Class 1, Basic Digital Building Blocks, that total 46.17\% modules. Classes 2 and 3, Combinational and Sequential Logic Designs, make up > 40\% of the \textit{VerilogDB} dataset. The remaining $\sim$12\% of the dataset is well distributed across the rest of the classes spanning from custom accelerators to security and crypto blocks. Provided the large number of data, 761 MB of Verilog data, oversampling minority classes or loss reweighting \cite{cui_loss} can be  implemented to prepare a more balanced sample for LLM fine-tuning. 

\subsubsection{Module Complexity.} 
\begin{figure}[h]
    \centering

    \begin{subfigure}[t]{0.45\textwidth}
        \centering
        \fbox{\includegraphics[width=\linewidth]{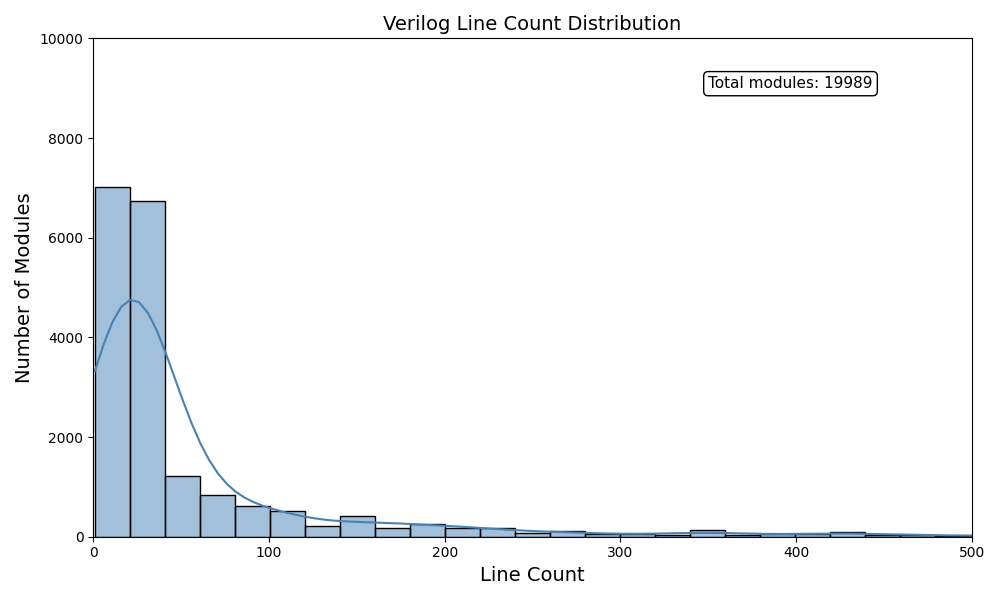}}
        \caption{Line counts.}
        \label{fig:line-count}
    \end{subfigure}
    \hfill
    \begin{subfigure}[t]{0.45\textwidth}
        \centering
        \fbox{\includegraphics[width=\linewidth]{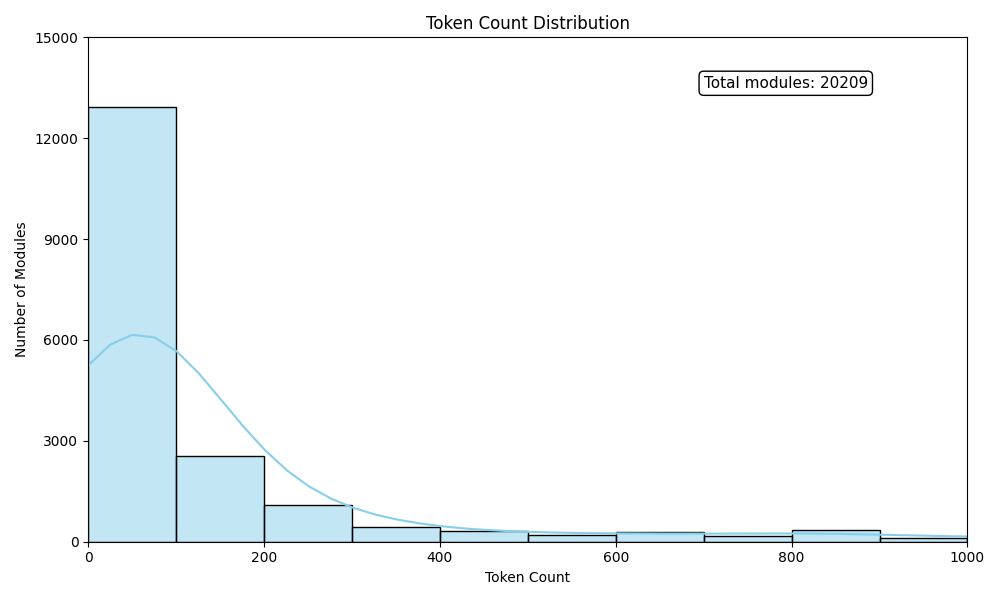}}
        \caption{Estimated token counts}
        \label{fig:token-count}
    \end{subfigure}

    \caption{Distributions of \textit{VerilogDB} modules by (a) functional classes, (b) line count, and (c) estimated token count.}
    \label{fig:verilog-distributions}
\end{figure}

\begin{figure}[h]
\centering
\fbox{\includegraphics[width=0.95\linewidth]{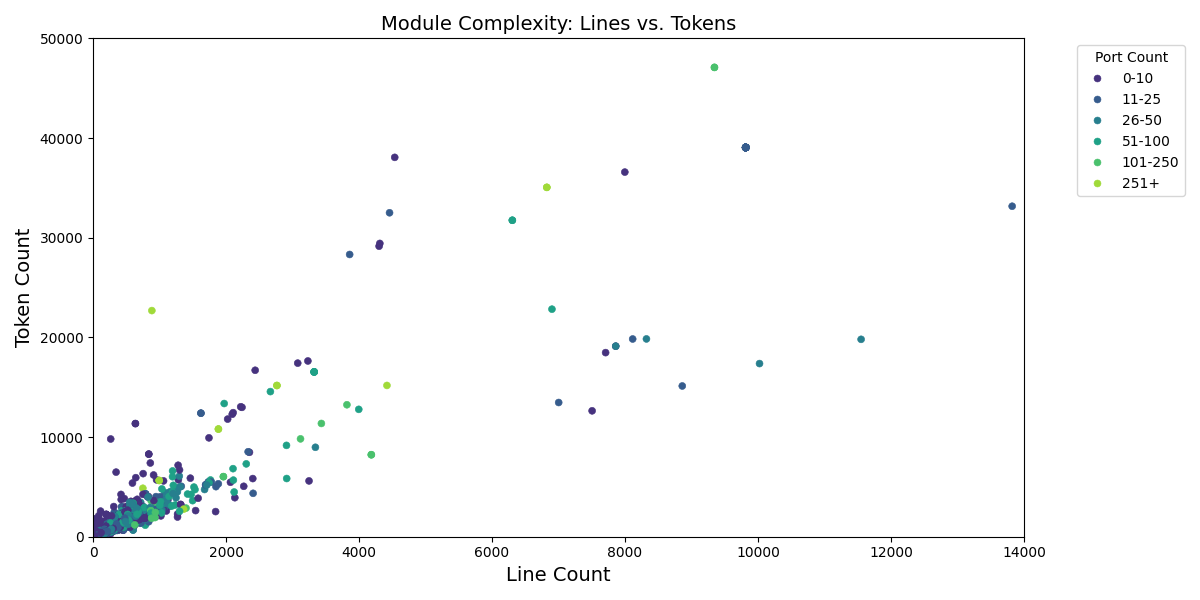}}
\vspace{-8pt}
\caption{Scatterplot showing of module complexity relating estimated token and line counts of the Verilog modules}
\vspace{6pt}
\label{fig:lines_vs_tokens}
\end{figure}

The dataset spans a wide spectrum of design scales. Some modules are compact, representing atomic components or student-written designs for educational use. Others are larger and structurally complex, composed of multiple submodules with layered hierarchy and interconnect logic. This range provides LLMs with examples for learning both isolated RTL constructs and compositional hardware behaviors. We utilize three key metrics to approximate a module's complexity: (1) estimated LLM token count to represent the module's size for downstream processing, (2) Verilog code line count (not including blank lines) to highlight how large the file was, and (3) the number of ports in the module. In general, more complex and larger modules would take up more lines of Verilog code and take more LLM tokens while typically having more I/O ports. We also note further analysis can be performed to estimate the module complexity like counting the number of \textit{always} blocks or \textit{case} statements per code or how large \textit{for generate} blocks will unravel, as for small line counts they can still relate to large complex designs. As seen in Figures~\ref{fig:verilog-distributions} and \ref{fig:lines_vs_tokens}, most modules in \textit{VerilogDB} are below 100 lines of Verilog code and are below 200 LLM tokens (15,475 modules). Also, most designs (17,715 modules) have ports from 0-10 with a port count mean $\mu_{pc}$ of 7.99 and median $M_{pc}$ (better adjusted for non-symmetrical data) of 4. The LLM token count has a mean $\mu_{tc}$ of 2126.34 and median $M_{tc}$ of 55, while the line count has a mean $\mu_{lc}$ of 700.16 and a median $M_{lc}$ of 28. The higher mean values compared to medians indicate that the dataset is right/positively skewed and not symmetrical, indicating the presence of fewer larger and more complex modules making the dataset more tailored for smaller to moderate designs, essential for the foundation of hierarchical designs. Sampling for LLM fine-tuning must take this into account to select a well-balanced distribution of the modules from the dataset to account for the skew. In future work, we seek to expand the dataset's range through data augmentation to address and mitigate biases.

\vspace{-10pt}

\subsubsection{Annotation and Comments.} 
\begin{figure}[h]
\centering
\fbox{\includegraphics[width=0.90\linewidth]{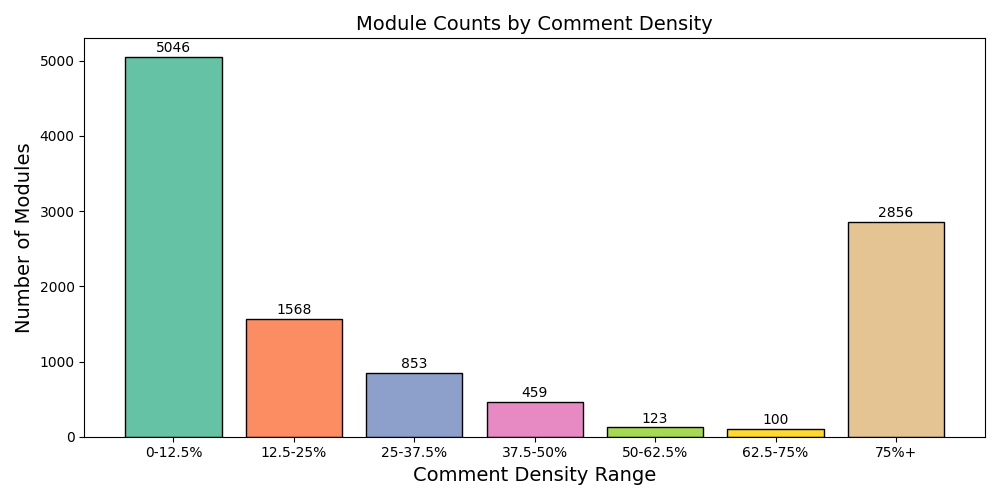}}
\vspace{-8pt}
\caption{Comment density module counts for 0-12.5\%, 12.5\%-25\%,25\%-37.5\%, 37.5\%-50\%, 50\%-62.5\%, 62.5\%-75\%, and >75\% buckets}
\vspace{6pt}
\label{fig:comments}
\end{figure}

A subset of the dataset, particularly those sourced from academic and textbook-derived content, contains meaningful inline comments and block-level summaries. When available, these annotations are preserved and extracted as auxiliary metadata for instruction generation and semantic indexing. While not uniformly present, their inclusion supports higher-quality prompt creation for instruction-tuned LLMs. In \textit{VerilogDB}, 53.9\% of the modules (11,009 modules) have comments. A larger subset of modules with comments would further enhance the quality understanding of the code. We seek to develop automated code commenting in future work to enhance the dataset and improve this component, as fine-tuning an LLM with Verilog code containing comments provides more context and can improve model ability~\cite{song2024codeneedscommentsenhancing,guo-etal-2022-unixcoder}. A critical metric to evaluate the quality of a codebase's commenting is code comment density \cite{code_comment}, defined as the percentage of comment lines of code in a codebase. We define comment density as the number of comment characters over the total characters of the Verilog code. Figure \ref{fig:comments} shows the distribution of modules based on ranges of comment densities. Of the 11,009 commented modules, 60.08\% of the modules have a code comment density $D_c$ where $0\% < D_c <= 25\%$. Modules where $25\% < D_c <= 50\%$ account for 11.92\% of the 11,009 commented modules. Lastly, 3,079 modules (27.97\% of commented codes and 15.1\% of total \textit{VerilogDB}) have $D_c > 50\%$. This reveals almost a third of the commented codes have more than half of the text coming from comments which are rich in context for LLM fine-tuning. Of the codes with comments, the mean $\mu_{Dc}$ is 1093.71\% shows the wide range of comment densities. The median comment density $M_{Dc}$ is 15.52\%, providing a more accurate view of the average $D_c$ that accounts for bias. 

\subsection{Discussion}
The construction and analysis of our \textit{VerilogDB} framework revealed a number of important benefits and challenges with VerilogDB's dataset curation, preprocessing workflows, and implications for downstream model development. These findings inform both our current pipeline design and future directions for scaling Verilog-focused LLM research. While our approach in the current framework focuses on a dataset for LLM fine-tuning, we also note that \textit{VerilogDB} can be leveraged for RAG-based approaches~\cite{gao2024retrievalaugmentedgenerationlargelanguage, koziolek_rag, wu2025retrievalaugmentedgenerationnaturallanguage,gupta2024comprehensivesurveyretrievalaugmentedgeneration} rather than fine-tuning; however, an additional vectorization stage would be necessary to produce the embeddings for queries. Table \ref{tab:verilogdb_insights_detailed} provides an overview of key insights from our approach and their implications on LLM fine-tuning. We discuss both the benefits of \textit{VerilogDB} compared to other works and the encountered challenges:

\subsubsection{Benefits} Our dataset's potential to augment automated hardware design is underscored through the following key points:

\begin{itemize}
    \item \textbf{A Rich and Diverse Dataset Enhances HDL Research.} As shown in Figures~\ref{fig:classes} and~\ref{fig:lines_vs_tokens}, the dataset captures a variety of module types, interface complexities, and design styles. Arising naturally from multi-source scraping, this variation increases LLM efficiency and helps avoid overfitting to any single design idiom. Other works use datasets with Verilog modules which have been intentionally restricted to stay within line count ranges, limiting the capabilities of trained LLMs and their generalization to varying module complexities and functionalities.
    
    \item \textbf{Rigorous Preprocessing Augments Quality.} Exact and near-duplicate modules were commonly found across GitHub repositories, especially in forked or template-based designs. Removing them reduced redundancy, improved design diversity, and minimized the risk of LLM overfitting to high-frequency samples. While syntax checking is a common filtering step, it proved insufficient on its own as many syntactically valid files failed synthesis due to unresolved submodules, macro dependencies, or simulation-only constructs. Our combination of syntax and synthesis checks, paired with metadata extraction, provides a more reliable foundation for model training. Other works include preprocessing in a more manual fashion with little automation or rely on LLMs to preprocess the data (syntax and synthesis checks) which are prone to hallucination and are not as robust as industry aligning compilation and logic synthesis tools.

    \item \textbf{Metadata and comment presence enable semantic alignment.} While the quantity and consistency of comments vary, their presence in over half of the dataset makes them valuable for prompt-response generation and instruction-tuning. This gives an opportunity for future work in automated comment enrichment. Other Verilog datasets for LLMs, however, widely do not consider comments as significant aspects to have span most of the dataset for context enrichment. In contrast, our approach stores all comment information as metadata.

    \item \textbf{LLMs must learn from real-world RTL noise.} Exposure to inconsistent formatting, incomplete modules, and stylistic variation in early preprocessing steps reinforces the need to train models not just on idealized textbook code, but also on the kinds of imperfect RTL encountered in practice. A small number of modules with excessive ports, unusual formatting, or partial declarations can serve as test cases to improve the model's robustness. Other works remove whitespaces of modules or alter the data to match consistent stylings/formats as a preprocessing step to lead to higher accuracy for modules with that same formatting; however, their capabilities to generate or understand Verilog with a wide diversity of stylings, expected in real-world applications, is yet to be seen.

\end{itemize}

\renewcommand{\arraystretch}{1.2} 
\begin{table*}[h]
\small
\centering
\caption{Summary of Verilog dataset sources and their contribution to LLM training.}
\vspace{-6pt}
\label{tab:verilogdb_insights_detailed}
\begin{tabularx}{\textwidth}{|>{\raggedright\arraybackslash}X 
                                |>{\raggedright\arraybackslash}X 
                                |>{\raggedright\arraybackslash}X 
                                |>{\raggedright\arraybackslash}X|}
\hline
\textbf{Insight Area} & \textbf{What We Observed} & \textbf{Concrete Examples / Evidence} & \textbf{Implications for LLM-Based RTL Learning} \\
\hline

\textbf{Preprocessing Effectiveness} & Major size reduction from raw to clean data. & $\sim$30 GB raw $\rightarrow$ 751 MB cleaned; 20,392 usable modules. & Filters out noise, reducing overfitting and improving convergence. \\
\hline

\textbf{Deduplication \& Validation} & High duplicate rate; multi-step filtering needed. & 12.47\% duplicates removed; syntax pass = 63.07\%, synthesis pass = 69.49\%. & Validations ensure training stability and correctness. \\
\hline

\textbf{Module Complexity \& Size} & Range from simple to deeply nested modules. & Most <200 tokens and 100 lines; some exceed 10k tokens. & Models learn from basic logic to complex SoC components. \\
\hline

\textbf{I/O Port Distribution} & Wide variation in port counts. & Most modules: 0–10 ports; some >500 (Figure~4b). & Exposes models to diverse interface widths. \\
\hline

\textbf{Comment Availability} & Many include inline documentation. & 53.9\% of modules (11,009) have comments; 28\% are over 50\% comments. & Boosts instruction tuning with rich prompt-target pairs. \\
\hline

\textbf{Metadata Extraction Success} & Structured formatting after synthesis pass. & 99.01\% converted to JSON with ports, tokens, source. & Enables prompt templates and supervision pipelines. \\
\hline

\textbf{Functional RTL Diversity} & Broad spectrum of design types. & Modules span 13 RTL classes (Figure~\ref{fig:classes}). & Supports generalization across hardware domains. \\
\hline

\textbf{Design Source Heterogeneity} & Sourced from varied ecosystems. & GitHub (varied), OpenCores (reusable IP), Academia (clean, instructional). & Enhances LLM robustness to style, naming, and structure. \\
\hline
\end{tabularx}
\end{table*}

\subsubsection{Challenges} In developing \textit{VerilogDB}, we encountered a variety of challenges:

\begin{itemize}
    \item \textbf{Data Collection and Heterogeneity:} Collecting 20,392 samples from diverse repositories posed significant challenges due to the heterogeneity in coding styles, lack of adequate documentation, and inconsistencies in module formatting. Variations in Verilog syntax and inconsistent naming conventions complicated metadata extraction where our implemented regex patterns had to cover all possible module definition and port definition constructs. Validation was required to resolve ambiguities, increasing preprocessing time.
    
    \item \textbf{Preprocessing Scalability:} The preprocessing pipeline, while effective, was resource-intensive for the large sample set including some very large modules. Synthesis validation, which required Yosys, was particularly time-consuming and took days to complete. Scaling to larger datasets (e.g. 100,000 samples) would necessitate hardware intervention to improve performance, such as better memory optimization, distributed computation for parallelization, and high-core count CPUs. Furthermore, description assignment relied on LLM-generated summaries from o3-mini, which also took days to complete and occasionally produced generic or inaccurate descriptions, requiring review for critical modules during the DB validation stage.
    \item \textbf{Future Data Relevance:} The rapid evolution of hardware design practices and the adoption of newer standards pose a challenge to \textit{VerilogDB's} long-term relevance. While the 20,392 samples capture a snapshot of open-source Verilog modules, they may be based on older coding standards, potentially limiting their applicability to modern design flows. Continuously updating the dataset to include contemporary designs requires ongoing data collection and preprocessing, which is resource-intensive. This challenge underscores the need for a sustainable maintenance strategy to keep the DB aligned with industry trends, hence why our VerilogDB emphasizes DB creation and maintanence as a core pillar of the work, which other works do not discuss. 

\end{itemize}

\subsubsection{Using VerilogDB in DeepC}
\textit{VerilogDB} also provides an effective DB for RAG-based RTL generation approaches. As mentioned above, current RTL generators use RAG or agentic approaches; however, none incorporate both to optimize performance. Our current work, DeepC, leverages \textit{VerilogDB} as the input for vector DB generation, enabling the retrieval system to have a large high-quality sample set to improve the model context. DeepC includes various agents for code correction, simulation/testbench generation and error checking, and code generation utilizing the vectorized \textit{VerilogDB} through RAG. With our method, agents can leverage state-of-the-art LLMs as model architectures improve over time. In addition, RAG allows for more up-to-date Verilog codes to be added to the vector DB enhancing relevancy and facilitating industry partners to add proprietary IP. DeepC will be released soon.

\section{Conclusion and Future Work} \label{section:conclusion}
As LLMs have seen rapid growth in usage across most industrial sectors, particularly in technology, they can improve the acceleration of the hardware design flow. In our work, we showcase \textit{VerilogDB}, a framework for developing a custom Verilog dataset for LLM fine-tuning for RTL code generation. We evaluated our three-part approach that involved data curation, preprocessing, and custom DB architecture and storage with a raw input dataset of $\sim$ 30 GB to produce more than 751 MB of high-quality Verilog data. This large dataset With \textit{VerilogDB}, we intend to expand the dataset and develop fine-tuning and RAG agentic approaches for HDL generation. We also seek to label the data by functionality, further enhancing module context for LLMs. Lastly, another critical goal for future work includes LLM secure-RTL generation for hardware IPs to meet evolving security specifications in a semiconductor space rife with hardware threats.

\newpage
\appendix
\section*{Appendix}
\label{appendixA}

\begin{longtable}{|p{0.25cm}|p{3cm}|p{10.5cm}|}
\caption{Class names and Definitions for VerilogDB Modules}
\label{tab:class_table} \\
\hline
\textbf{\#} & \textbf{Class Name} & \textbf{Example Components} \\
\hline
\endfirsthead
\hline
\textbf{\#} & \textbf{Class Name} & \textbf{Example Components} \\
\hline
\endhead

1 & Basic Digital Building Blocks & Multiplexers, Demultiplexers, Decoders, Encoders, Half Adders, Full Adders, Subtractors, Comparators, Parity Generators, Priority Encoders, Registers (D, T, JK, SR), Shift Registers (SISO, SIPO, PISO, PIPO), Binary Counters, Ring Counters, Johnson Counters \\
\hline

2 & Combinational Logic Designs & Combinational ALUs (add, subtract, AND, OR, XOR, NOT, shift), Logic Gates (AND, OR, NOT, NAND, NOR, XOR, XNOR), Bit-slice architectures, Barrel Shifters, Carry Lookahead Adders, Booth Multipliers, Wallace Tree Multipliers, Binary to Gray and Gray to Binary Converters \\
\hline

3 & Sequential Logic Designs & Finite State Machines (Moore, Mealy), Pipelined Data Paths (IF, ID, EX, MEM, WB stages), Clock Dividers, Timers, Watchdogs, Register Files with Read/Write Ports, Sequential ALUs with pipelined operation, Sequence Detectors, Pulse Stretchers, Edge Detectors \\
\hline

4 & Memory Components & SRAM Controllers, DRAM Controllers (LPDDR4, DDR3/4), ROMs (fixed, OTP), EEPROM controllers, FIFOs (synchronous/asynchronous), Dual-Port RAMs, CAMs (Content Addressable Memories), Cache Controllers (write-through, write-back), Memory Protection Units (MPUs), Page Table Walkers \\
\hline

5 & Communication Interfaces & UART, SPI (Full-duplex, Quad SPI), I2C (Standard, Fast, High-Speed), CAN, USB 2.0/3.0, PCIe PHY/MAC, AXI Interconnects, AHB Masters/Slaves, APB bridges, DMA Engines (single and burst transfer), Ethernet MAC/PHY (10/100/1000 Mbps), JTAG TAP Controllers \\
\hline

6 & Security and Crypto Blocks & AES (128/256-bit), RSA Encryption/Decryption, SHA-1/SHA-2 Hash Engines, HMAC, True Random Number Generators (TRNGs), Pseudo-Random Number Generators (PRNGs), Secure Boot Engines, Key Management Units, Access Control Checkers, Secure Debug Locks \\
\hline

7 & SoC Integration Components & Bus Arbiters (round-robin, priority), Crossbar Switches, Address Decoders, Interrupt Controllers (e.g., PLIC), Reset Controllers, Clock Domain Crossing (CDC) Synchronizers, Power Management Units, Voltage Regulators, System Control Registers, Boot ROM Mappers \\
\hline

8 & Processor and Core Subsystems & Custom RISC-V / ARM cores, Microcode Engines, Instruction Fetch/Decode Units, Register Renaming Units, Reorder Buffers, Branch Target Buffers, Return Address Stacks, TLBs (Translation Lookaside Buffers), L1/L2/L3 Cache Hierarchies, MMUs (Memory Management Units) \\
\hline

9 & Peripheral Interfaces & GPIOs (bidirectional, open-drain), PWM Generators, ADC Interface Controllers, DAC Control Logic, Touchscreen Interfaces (resistive/capacitive), Camera Sensor Interfaces (MIPI CSI), Display Interfaces (HDMI, VGA, LVDS), SD/MMC Card Interfaces, Audio Codecs (I2S, AC'97) \\
\hline

10 & Verification-Only Designs & Basic Testbenches, Coverage Monitors (functional/code), Scoreboards for result checking, Bus Functional Models (BFMs), Interface Monitors, Checkers for protocol compliance, Reference Models \\
\hline

11 & Custom Accelerators and IP Blocks & ML Accelerators (Matrix Multipliers, Softmax, Activation Units), DSP Units (FIR, IIR, FFT, DCT), Video Codecs (H.264, H.265), Neural Network Inference Cores, JPEG Compression/Decompression, Hardware Sorting Units, CRC Generators, Hardware Schedulers \\
\hline

12 & Clock and Power Management Units & Clock Gating Units, PLL/DLL Interfaces, Frequency Dividers, Dynamic Voltage/Frequency Scaling Units, Power Gating Controllers, Wakeup Controllers, Retention Register Banks, Clock Mux/Demux Units \\
\hline

13 & Error Handling and Fault Tolerance Modules & ECC Encoders/Decoders (SECDED, BCH, Reed-Solomon), Watchdog Timers, Parity Generators/Checkers, Redundancy Logic (TMR – Triple Modular Redundancy), Fault Injection and Detection Blocks, Error Logging Units \\
\hline

\end{longtable}

\newpage
\begin{acks}
The author(s) would like to thank Shams Tarek for valuable assistance with the classification portion of this work.
\end{acks}

\bibliographystyle{ACM-Reference-Format}










\end{document}